\documentclass[10pt,fleqn]{article}
\makeatletter
  \def\eqalign#1{\null\,\vcenter{\openup\jot\m@th
    \ialign{\strut\hfil$\displaystyle{##}$&$\displaystyle{{}##}$\hfil
      \crcr#1\crcr}}\,}
\def\cmeqalign#1{\null\,\vcenter{\openup\jot\m@th
  \ialign{\strut\hfil$\displaystyle{##}$&&\hfil$\displaystyle{{}##}$\hfil
      \crcr#1\crcr}}\,}
\makeatother
\let\Message=\typeout
\def\oversymbol#1#2{\vbox{\ialign{##\crcr \hfil$#1$\hfil\crcr
   \noalign{\kern1pt\nointerlineskip}%
   \hbox{$\hfil\displaystyle#2\hfil$}\crcr}}}
\def\overeq#1{\oversymbol{\scriptstyle\kern.5pt =}{#1}}
\def\dualp#1{{}^{\ast_{(\hbox{$\scriptstyle #1$})}} \kern-1pt}

\def\bivec#1{\vbox{\ialign{##\crcr $\leftrightarrow$\crcr\noalign{
   \kern-1pt \nointerlineskip}$\hfil\displaystyle{#1}\hfil$\crcr}}}
  
   \def\pub(#1){(#1)}
\def\be{\begin{equation}}
\def\ee{\end{equation}}
\def\Int{\int\limits}

\def\defin{\stackrel{\rm def}{=}\ }
\def\vainf#1{\raisebox{-1.5ex}{$\stackrel {\displaystyle {\longrightarrow } }
        {\scriptscriptstyle{#1\rightarrow\infty }} $}}

\def\llongrightarrow{{-\!\!\!-\!\!\!-\!\!\!\!\longrightarrow}}

\def\minord{\raisebox{-.8ex}{$\ \stackrel{<}{\sim}\ $} }

\def\p{{{\bf p}}}
\def\q{{{\bf q}}}

\def\v{{{\bf v}}}

\def\x{{{\bf x}}}

\def\z{{{\bf z}}}
\def\D{{{\bf D}}}

\def\bfxi{{\mbox{\boldmath$\xi$}}}

\def\Ucal{{\cal U}}
\def\Tcal{{\cal T}}

\def\Ncal{{\cal N}}
\def\Hcal{{\cal H}}
\def\Rcal{{\cal R}}

\def\Rc0{{\Rcal^{\!\!{}^{\scriptscriptstyle[0]}}}\!}
\def\bfHcal{{\mbox{\boldmath$\Hcal$}}}

\def\s|{\left|\!\left|}
\def\d|{\right|\!\right|}

\def\qquad{\  \  \  \  }
\def\jlc{{{\sf JLC~}}}

\def\BIX{{\sf BIX}}

\def\eg{{\tt e.g.}}
\def\ie{{\rm i.e.}}
\def\lcn{{\sf LCN}}

\def\DS{{\sf DS}}
\def\GDA{{\sf GDA}}

\def\rhs{{r.h.s.~}}

\def\TqM{{{\sf T}_\q{\sf M}}}
\begin{document}
\begin{titlepage}
\baselineskip20pt
\setcounter{page}{1}
\begin{center} 
{\LARGE 
{\bf {Geometry and Chaos\\
on Riemann and Finsler Manifolds\\}}
}
\vskip3.truecm 
{\Large 
Maria DI BARI$^{1\ast}$\\
Piero CIPRIANI$^{1,2\dagger}$\\
} 
\vskip.8truecm 
{\large
$^1${Dipartimento di Fisica {\it "E. Amaldi"}, Universit\`a 
{\it "Roma Tre"},\\
Via della Vasca Navale, 84 \ -- \ {\bf 00146 ROMA, Italia}}\\
$^2$  {\sf I.N.F.M. - sezione di ROMA.}\\
}
\end{center}
\vskip3.3truecm 
{\large 
{\bf Submitted to}: Planetary and Space Science\\
}
\vskip2truecm
\null\noindent
{\large 
{\bf Send proofs to:} Piero CIPRIANI (address above)
}\\
{\large 
{\bf Send offprint request to:} Piero CIPRIANI (address above)
}\\
\vfill\baselineskip12pt
\noindent
\hrulefill\hspace{9truecm}\ \\
$^\ast$    {\small{\tt e-mail: DIBARI@VXRMG9.ICRA.IT}}\\
$^\dagger$ {\small{\tt e-mail: CIPRIANI@AMALDI.FIS.UNIROMA3.IT}}\\
\end{titlepage}
\Message{Titlepage:}
\setcounter{footnote}{0}
\setcounter{page}{2}
\begin{abstract}
\baselineskip14pt
In this paper we discuss some general aspects of the so-called
{\sl geometrodynamical approach} ({\sf GDA}) to Chaos and present
some results obtained within this framework. We firstly
derive a {\sl na\"\i ve} and yet general
 geometrization procedure, and then
 specialize the discussion to the descriptions of motion within 
the frameworks of two among the most representative 
implementations of the approach, namely the Jacobi and the 
Finsler {\sl geometrodynamics}.
In order to support the claim that the \GDA~isn't simply
a mere re-transcription of the usual dynamics, but instead
can give various hints on the understanding of the qualitative
behaviour of dynamical systems (\DS's), we then
compare, from a formal point of view, the tools used
within the usual framework of Hamiltonian dynamics to detect the
presence of Chaos with the corresponding ones used within the
{\sf GDA}, \ie, the tangent dynamics
and the geodesic deviation equations, respectively,
pointing out their general inequivalence.
Moreover, to go ahead
the mathematical analysis and to highlight both the peculiarities
of the methods and the analogies between them,
we work out two concrete 
applications to the study of very different, yet typical in distinct
contexts, dynamical systems. 
The first is the well-known H\'enon-Heiles Hamiltonian, which allows
to exploit how the Finsler \GDA~is well suited not only for
testing the dynamical behaviour of systems with few degrees of freedom,
but even to get deeper insights on the sources of instability.
We show the effectiveness of the {\sf GDA}, both in recovering fully
satisfactory agreement with the most well established outcomes and also
in helping the understanding of the sources of Chaos.\ 
Then, in order to point out the general applicability of the method,
we present the results obtained from the geometrical description 
of a General Relativistic \DS, whose peculiarity is 
well known, as its very nature has been debated since long time,
namely the Bianchi IX ({\sf BIX}) cosmological model.
Using the Finsler \GDA, we obtain results with a {\sl built-in
invariance}, which give evidence to the non chaotic behaviour 
of this system, excluding any global exponential instability 
in the evolution of the geodesic deviation. 
\end{abstract}
\newpage
\baselineskip16pt
\section*{Introduction.}
The investigations on the occurrence of regular and chaotic behaviour
in $N$ dimensional dynamical systems are performed with a 
variety of methods and mathematical tools.
Recently, this ensemble widened with the inclusion
of the {\sl geometrodynamical approach} ({\sf GDA}), 
\eg, \cite{Marco93,CPTD}.
As other {\sl new approaches}, this tool has been suggested and 
applied to the study of 
stability properties of general dynamical systems, 
in the hope to bring the phenomenological analysis of their
possibly chaotic behaviour back to an inquiry directed towards
an {\sl explanation}, at least qualitative, of the mechanisms
responsible for the onset of Chaos. Within the framework of the
geometrical picture, this {\sl explanation} was sought through
a possible link between a change in the curvature properties of 
the underlying manifold and a modification of the qualitative 
dynamical behaviour of the system.\\ 
Within the Hamiltonian approach, the ingredients needed to
make Chaos lie basically on the presence of {\sl stretching and folding}
of dynamical trajectories; \ie, in the existence of a {\sl strong
dependence on initial conditions}, which, together
with a bound on the extension of the phase space, yields to a substantial
unpredictability on the long time evolution of a system.
Usually, the strong dependence is detected looking at the
occurrence of an exponentially fast increase of the {\sl separation} 
between initially arbitrarily close trajectories. To have {\sl true Chaos},
this last property must be however supplemented by the {\sl compactness} 
of the ambient space where dynamical trajectories live, this simply
in order to discard trivial exponential growths due to the unboundedness
of the {\sl volume} at disposal of the \DS. Stated otherwise, the
folding is necessary in order to have a dynamics actually able to
{\sl mix} the trajectories, making practically impossible, after a finite
interval of time, to discriminate between trajectories which were
very nearby each other at the initial time. When the space isn't
compact, even in presence of strong dependence on initial
conditions, it could be possible,
in some instances (though not always), to distinguish among
different trajectories originating within a small distance and
then evolved subject to exponential instability.\\
In the geometric description of dynamics, the recipe to find
Chaos is essentially the same, with some generally speaking
minor differences, which nevertheless prove sometimes to be very
relevant for the understanding of the qualitative behaviour of the system.
When the geometrization procedure is accomplished, the study of
dynamical trajectories is  brought back to the analysis of a geodesic
flow on a suitable manifold {\sf M}. In order to define Chaos, {\sf M}
should be {\sl compact}, and geodesics on it have to deviate 
{\sl exponentially fast}. As we will show below, now the point
representative of the state of the system will move along
geodesics with {\sl unit speed} and the dynamics itself determines the
{\sl natural way} to measure distances on {\sf M} (or, rather, 
on ${\sf T}_{\{\q\}}{\sf M}$).
\ Though there exist
{\sl many} geometrizations, \cite{CPTD,MTTD}, in the following we will
exploit mostly two of them, namely, the Jacobi and Finsler ones,
 essentially equivalent for most \DS's with many degrees of freedom, 
but whose differences result very enlightening
on the sources of instability in the case of few dimensional systems.\\
The \GDA~can be adopted to cope with very general \DS's, \eg,
\cite{PRE97a}; however, in the present paper,
we will confine to discuss {\sl only} that (still very large) subset
of systems for which a comparison between the usual Hamiltonian approach
and the two geometrical ones here discussed can be carried out
without troubles.
So, here and in the sequel, we will concentrate on 
$N$-dimensional conservative dynamical system. We further suppose,
 only to simplify the notation, that
a choice of coordinates, $\x=\{x^i\}$, exists such that the equations 
of motion can be put in the {\sl newtonian form}:
\be
\ddot x^i+\Ucal_{,i} = 0 \ \ \qquad\ \ i=(1,\dots,N)  \ ,
\label{tmotion}
\ee
where $\Ucal=\Ucal(\x)$ is the potential, we denoted by a dot the derivative
with respect to {\sl newtonian} time and indicated the partial derivative
with respect to one of the coordinates as
\[
\Ucal_{,i}={\displaystyle {\partial \Ucal\over \partial x^i}}\ .
\]
The analysis of the stability of the dynamics given by eq.(\ref{tmotion})
is {\sl usually} performed studying the
evolution of two initially close trajectories, starting
at $t=0$ from two very {\sl nearby} points 
${\bf P}_0$ and $\overline{\bf P}_0$, \ie,
with very similar initial conditions
\[
{\bf P}_0 = [\x(0)\, ,\,\dot{\x}(0)] \qquad\ ;\qquad\  
\overline{\bf P}_0 = [\x(0)+\bfxi(0)\, ,\,\dot{\x}(0)+\dot{\bfxi}(0)]\ ,
\]
where the {\sl disturbance vector}, $\D=\{D^A\},\ (A=1,\ldots,2N)$,
is defined through 
\be
\D =(\bfxi,\dot{\bfxi})\qquad \Longleftrightarrow \qquad
\{D^A\} = (D^1,\ldots,D^N,D^{N+1},\ldots,D^{2N}) 
        = (\xi^1,\ldots,\xi^N,\dot\xi^1,\ldots,\dot\xi^N)\ ,
\ee 
\ie, $D^i= \xi^i\ ;\ D^{N+i} = \dot\xi^i\ , (i=1,\ldots,N)$.
This vector represents
the uncertainty in the initial conditions, and connect the two points 
in the {\sl space of states} of the system. With the choice of
coordinates assumed above, it gives
also the separation in phase space.
The evolution of the {\sl small} perturbation vector is governed
by the tangent dynamics equations, \cite{BGS76,BGS80},
which read
\be
\ddot\xi^i +\Ucal_{,ij} \ \xi^j = 0 \label{tvar} 
\ \ \qquad\ \ i=(1,\dots,N)  \ ,
\ee
where clearly
\[
\Ucal_{,ij}={\displaystyle {\partial^2 \Ucal\over 
\partial x^i \partial x^j}}\ ,
\]
and the summation convention in the products is adopted.\\
A quantitative indicator of the presence of Chaos in the dynamics is
extracted from the solutions of eq.(\ref{tvar}), measuring the 
possible exponential growth of the {\sl distance} between trajectories,
assumed to be represented by the norm, $D(t)= \|\D(t)\|$, of the separation 
vector in phase space.
The (maximal) Lyapunov characteristic number ({\sf LCN}) is indeed defined
as the asymptotic quantity, \cite{BGS76}:
\be
LCN= \lim_{t \rightarrow\infty}
 \ \lim_{D(0)\rightarrow 0}\ \ 
		      \left[ {\displaystyle {{1\over t} \ln {{D(t)}
		       \over {D(0)}} } }\right] \quad ,
\label{LCN}
\ee
where a suitable norm of a $2N$-dimensional vector in phase space 
should have been
defined, with the introduction of a $(2N\times 2N)$ {\sl metric}
$\Upsilon_{AB}$: 
\[
\|\D(t)\|^2 = \Upsilon_{AB} D^A D^B
\]
Up to this point, the choice of the metric is largely arbitrary,
as the dynamics does not give any prescription on it,
not to mention that, moreover, the phase space do not possess
a truly metric structure. This indefiniteness is usually exploited
in its full freedom, imposing an Euclidean metric to the full
$2N$-dimensional space, distinguishing also coordinates and
momenta spaces, \ie, putting
\be
D^2(t)=\|\D(t)\|^2 =
 \eta_{ij} \left( \xi^i \xi^j + \dot\xi^i \dot\xi^j \right)\ .
\ee
A \DS~is said to be chaotic in a region ${\cal W}$ of its phase space if
${\cal W}$ is compact, orbits starting inside ${\cal W}$ remains
into ${\cal W}$, and almost all orbits starting in its interior are 
characterized by a positive value of the maximal \lcn, for almost any 
choice of the initial perturbation $\D(0)$.
\section*{Dynamics and Geometry}
Although the geometrization procedure can be carried out elegantly and
in a full generality, substantially starting from different formulations
of the {\sl least action principle}, \cite{SyngeSchild,CPTD,Marco93},
we will show how a metric is naturally introduced looking at the
dynamics itself. 
Let's suppose that we are dealing with a \DS~as that given by 
eq.(\ref{tmotion}) and we perform a change of the time variable from the 
{\sl newtonian time}, $t$, to a new variable, $s$, defined through a
relation which in general depends, on time, coordinates and velocities:
\be
ds = f(t,x^i,\dot x^i) dt \ . \label{ds}
\ee
As a result, the equations of motion (\ref{tmotion}) become
\be
{d^2 x^i\over ds^2} + {1\over f}{df\over ds} {dx^i\over ds} + 
{1\over f^2} U_{,i} =0 \ . \label{Geo}
\ee
Depending on the choice of the function $f$, it can happens that the
{\sl new} equations of motion can still be derived from a variational
principle, $\delta {\cal A} = 0$, where
\be
{\cal A} = \int_{P_1}^{P_2} ds = \int_{P_1}^{P_2} f(t,x^i,\dot x^i) dt
\ee
and the integral is taken between fixed endpoints:
$P_1=P_1(t_1)=P_1(s_1)$ and $P_2=P_2(t_2)=P_2(s_2)$.
Provided that $f$ satisfies suitable conditions, 
 ${\cal A}$  can be thought as the lenght of a geodesic 
joining $P_1$ and $P_2$ and satisfying equation (\ref{Geo}); in this
case, $f$ also defines the lenght of the vector ${\bf \dot x}$.\\
When $f$ depends explicitly on time $t$, in order to complete the procedure,
it is necessary to include also the time as a coordinate. To this goal, we
introduce a further {\sl time} parameter $w$, and 
equation (\ref{ds}) becomes
\be
ds = f \left(t, {\bf x},{d{\bf x}\over dt}\right) dt = 
f \left(t, {\bf x},{d{\bf x}/dw\over dt/dw}\right){dt\over dw} dw \ .
\ee
If we put $x^0=t$ and indicate with a prime the derivation with
respect to the {\sl new time}, $w$, equation (\ref{ds}) reads
\be
ds = F(x^\alpha,x^{\prime\alpha})dw \qquad \alpha=0,1,\dots,N \ ,
\ee
where 
\be
F(x^\alpha,x^{\prime\alpha}) = f(x^\alpha, x^{\prime\alpha}) x^{\prime 0}\ .
\label{Fdif}
\ee
The number of equations of motion is now $(N+1)$. Indeed, from the variational
principle it follows 
\be
{d\over dw}\left[{\partial F(x^\alpha,x^{\prime\alpha}) \over 
\partial x^{\prime \beta}}
\right] - {\partial F(x^\alpha,x^{\prime\alpha}) \over \partial x^{ \beta}} = 0
\qquad \beta = (0,1,\dots,N) \ ,
\ee
and the additional equation, \ie, that with $\beta = 0$,
 gives the {\it conservation of energy}.
So when the time reparametrization involves an explicit dependence
on the {\sl old} time $t$, 
the geometrization procedure must be carried out using
the function of $(2N+2)$ variables $F(x^\alpha,x^{\prime\alpha})$,
instead of the original one $f(t,{\bf x},{\bf\dot x})$, depending
on only $(2N+1)$ variables.\\
In order to set a self consistent geometrodynamics, the 
function\footnote{Or either the function $f(\x,\dot\x)$, when the 
transformation do not involves explicitly the time $t$.}  
$F(x^\alpha,x^{\prime\alpha})$ must fulfil the following conditions:
\begin{description}
\item{a)}
$F(x^\alpha, x^{\prime\alpha})$ must be sign definite, e.g. positive
\be
F(x^\alpha, x^{\prime\alpha}) > 0 \qquad \forall \ x^{\prime\alpha}\not= 0 \ ,
\ee
with $F(x^\alpha, x^{\prime\alpha}) = 0\ $ if and only if
 $\ x^{\prime\alpha}=0\ \ \forall\ \alpha=(0,1,\ldots,N)$.
\item{b)}
$F(x^\alpha, x^{\prime\alpha})$ must be a positively homogeneous 
function of first degree in the $x^{\prime\alpha}$, i.e.
\be
F(x^\alpha,k x^{\prime\alpha})=k F(x^\alpha, x^{\prime\alpha}) \ , \quad
\forall\ k>0 \ ;
\label{con13}
\ee
in such a way $F\,dw$ is apparently invariant for reparametrization:
\be
F\left(x^\alpha,{dx^\alpha\over dw_1}\right) dw_1 \equiv
F\left(x^\alpha,{dx^\alpha\over dw_2}\right) dw_2 \ . \label{F_invar}
\ee 
\item{c)}
Finally,
\be
det {{\partial^2 F^2(x^\alpha, x^{\prime\alpha})}\over{\partial 
x^{\prime \beta}\partial x^{\prime \gamma}}} \neq 0 \ .
\ee
\end{description}
From the definition, eq.(\ref{Fdif}), the {\sl extended} function $F$
turns out to satisfy always condition b) above, while this is in general
not true for the function $f$.
So, if a built-in invariance for reparametrization, eq.(\ref{F_invar}),
is sought, even if the trasformation 
$t\rightarrow s$ does not explicitly depends on $t$, it is 
sometimes however necessary to introduce the function $F$.\\
If the conditions a), b) and c) are fulfilled, it is possible to exploit
the transormation obtained, in order to derive a metric {\sf g} on the 
manifold {\sf M}. It can be shown that the quantity
\be
g_{\alpha\beta}(x^\alpha, x^{\prime\alpha}) \defin {1\over 2} 
{{\partial^2 F^2(x^\alpha, x^{\prime\alpha})}
\over{\partial x^{\prime \beta}\partial x^{\prime \gamma}}} \ ,
\label{metrica}
\ee
meets all the requirements demanded to a metric.
By the properties of $F$, it easily follows, \cite{Rund,MTTD}, that the line 
element of eq.(\ref{metrica}) can be written as
\be
ds^2=F^2(x^\alpha, x^{\prime\alpha}) dw^2 = 
g_{\beta\gamma}(x^\alpha, x^{\prime\alpha})dx^{\beta}dx^{\gamma}
\ ;
\ee
and the equations of motion turn out to be the geodesic equations on
({\sf M,g}), which read:
\be
{d^2 x^\alpha\over ds^2} + \Gamma^\alpha_{\beta \eta} {dx^\beta\over ds} 
{dx^\eta\over ds} = 0 \ , \label{Geog}
\ee
where
\be
\Gamma^\alpha_{\beta \eta}(x^\rho,{x^{\prime \rho}}) =  
{{1}\over{2}}g^{\delta\alpha}{\left(
{{\partial g_{\delta\beta}}\over{\partial x^\eta}} + 
{{\partial g_{\delta\eta}}\over{\partial x^\beta}}
- {{\partial g_{\eta\beta}}\over{\partial x^\delta}} 
\right)}
\ee
are the connections of the metric, which, in the general case, depend also
on the {\sl tangent vectors} $x^{\prime\alpha}$.
If we further add to the conditions a), b) and c) the following
requirement: 
\be
{{\partial^2 F^2(x^\alpha, x^{\prime\alpha})}\over{\partial x^{\prime \beta}
\partial x^{\prime \gamma}}} v^\beta v^\gamma > 0 \qquad \forall \ 
v^\alpha\not= 0 \ ,
\ee
then we will deal with a positive definite metric.\\
As the dynamics determines the natural metric by itself, we see why
it is not necessary to {\sl impose} an  {\it a priori} metric, 
as has been done to measure the magnitude, $D(t)$, of the perturbation
vector obtained solving eq.(\ref{tvar}) along the trajectories, solutions
of eq.(\ref{tmotion}).
Within the framework of any geometrization, the norm of any vector
of the tangent space, $\v\in\TqM$, to the manifold on a point 
$\q\in {\sf M}$, simply reads:
\be
v^2\defin \|\v\|^2 =g_{\alpha\beta}(x^\alpha, x^{\prime\alpha}) 
v^\alpha v^\beta \ ; 
\ee
and we remark that in the most general case up to now discussed, the
metric {\sf g} as well the norm of the vector $\v$, depend not only
on the point $x^\alpha$ of the manifold, but also on the tangent vector
$x^{\prime\alpha}$.\\
The metric $g_{\alpha\beta}(x^\alpha, x^{\prime\alpha})$ is a {\sl Finsler 
metric}
and its geometrical features are widely developed (see \cite{Rund} and 
references therein). In \cite{MTTD} and in \cite{PRE97a}, the general mathematical 
formalism is derived and
are also presented some applications to several \DS's.\ 
At this point we observe that, in the case the metric do not actually
depends on the tangent vectors, \ie, when
\be
{\displaystyle F^2(x^\alpha, x^{\prime\alpha}) = 
g_{\alpha\beta}(x^\eta){dx^\alpha\over dw}{ dx^\beta\over dw}}\ ,
\ee 
then the manifold reduces to a Riemannian one.\\
We summarize our discussion emphasizing how, under very mild assumptions,
the geodesic equations can be recovered simply re-interpreting the
equations of motion of a \DS, writing down them in terms of a {\sl new} 
time parameter, and looking for a suitable transformation which make
the new parameter {\sl one} action of the system.\\

We now specialize our presentation to the two metrics mentioned in the 
Introduction, which can be obtained from the general discussion just
presented, by particular choices of the function $F$.
 The first one is the well known {\sl Jacobi} metric, which is a very
special case in that it is defined on a Riemannian manifold; whereas
the other {\sl dynamical} manifold we will explore is a {\sl pure Finsler} 
metric, as described above, obtained, as we will see, by a very 
{\sl familiar} function $f$.\\
Though simpler, the Jacobi metric has shown its reliability in describing
the main qualitative and quantitative properties of typical many degrees 
of freedom \DS's (see \cite{Aquila1} and references therein for a more detailed 
account) and has been recently applied with partial success 
also to the investigations of dynamical behaviour of few dimensional systems.
Indeed our attempt towards a generalization of the riemannian approach
has been motivated just because of the limitations encountered by the
Jacobi \GDA~in coping with two degrees of freedom systems or {\sl non
standard} ones, as those of General Relativitistic origin, which are
characterized by a very peculiar kinetic energy form.
In addition, it applies only to conservative systems\footnote{Being 
however relatively straightforward the inclusion of explicitly time
dependent potentials.}, and it is then not suitable for the study of 
lagrangian systems with gyroscopic terms, as those arising either when a non
inertial reference frame is chosen, \eg, the restricted three body problem,
\cite{MTTD,PRE97a}, or when electromagnetic interactions are taken
into account.
The \GDA~in the framework of {\sl Finsler} manifolds turns out to be
 much more general than the riemannian one, not only since it allows 
to describe a wider class of \DS's, but also because it gives more reliable
results in all the situations the Riemannian approach get in troubles,
as for few dimensional \DS's, {\sl at the border} between regularity and
Chaos, when the quasi integrable features of the dynamics, together with
the small number of degrees of freedom, causes the results obtained
within the riemannian approach to become noisy and less decipherable.
Some of the results obtained for two and three degrees of freedom systems
will be presented in the next sections, whereas a more detailed account, 
for both low and high dimensional \DS's, will appear elsewhere, 
\cite{CPMT_HH,MTCP_BIX} (see also \cite{Aquila1}). Some preliminary results
have been accounted for in \cite{PRE97a,Sigrav96}.

\subsection*{Riemann Manifold: Jacobi Metric}
Exploiting  the least action principle for holonomic 
conservative systems, in the formulation given by
Maupertuis, the geometrization procedure is accomplished, leading to the
well known Jacobi metric, \eg, \cite{ArnoldMC}.
The relation between the geodesic parameter and the 
{\sl newtonian time} is given by the transformation's law
\be
ds_J = \sqrt{E-\Ucal({\bf x})}\sqrt{2\Tcal}dt\ ,
\ee
where $E$ is the total energy, $\Ucal({\bf x})\,$ and
${\displaystyle \Tcal = {1\over 2} a_{ij}\dot x^i \dot x^j}$ 
are  the potential and kinetic energies, respectively.
It is easy to see that the function $f_J(\x,\dot\x)$
\be
f_J (\x,\dot\x) =  \sqrt{E-\Ucal({\bf x})}\sqrt{2\Tcal}
\ee
satisfies all the conditions a), b) and c), it is not necessary
to introduce the function $F$, and then the Jacobi metric
is $N$ dimensional and reads
\be
g_{ij}=\left[ E-\Ucal({\bf x})\right] a_{ij} \qquad i,j=(1,\dots,N)\ .
\ee
\subsection*{Finsler Metric}
If we want to be able to cope with the most general, possibly peculiar, \DS, 
we are naturally led to the introduction of the Finsler metric. While its 
derivation can be also obtained in a {\sl customary} fashion, \cite{Rund},
we will keep the spirit of the general presentation above: 
starting from the Hamilton least-action principle, 
a new {\sl evolution parameter} is defined through the 
Lagrangian of the system, $L$
\be
ds_F = L(t,\x,\dot\x) dt \ .
\ee
In the formalism of equations (\ref{ds}) and following, we have now that
the transformation law could in principle involve also the time $t$,
\[
f_F = L(t,{\bf x},{\bf \dot x})\ .
\]
\noindent In addition to the possible explicit time dependence, 
the requirement to fulfil condition b), eq.(\ref{con13}),
enforces to adopt the more general transformation law, introducing a
new parameter $w$ and defining the function of $(2N+2)$ arguments,
\be
F (x^\alpha, x^{\prime\alpha}) = 
L \left(t,{\bf x}, {{\bf x^\prime }\over x^{\prime 0}}\right) 
x^{\prime 0} \qquad \ \alpha =(0,1,\dots,N) \ .
\ee
If we are dealing with a conservative system whose lagrangian
function is
\be
L = {1\over 2} a_{ij}\dot x^i \dot x^j - \Ucal(x^i) \ ,
\ee
then the function $F$ reads
\be
F = {1\over 2 x^{\prime 0}} a_{ij} x^{\prime i}x^{\prime j} - 
\Ucal(x^i)x^{\prime 0} \ .
\ee 
It is an easy, though tedious, task to derive the metric, \cite{MTTD},
which is now $(N+1)$-dimensional, and depends, as remarked previously,
also on the {\sl velocities}, $x^{\prime\alpha}$:
\[
g_{00} = {1 \over 2} 
{\partial^2 F^2 \over \partial^2 (x^{\prime}{}^0)^2} 
                 =  3 \Tcal^2 + \Ucal^2 \ ,
\]
\be
g_{0a} = {1 \over 2}
{\partial^2 F^2 \over \partial x^{\prime}{}^0 \partial
x^{\prime}{}^a} = {\displaystyle  - 2 \Tcal a_{ab} {{x^{\prime}{}^b} \over
                   {x^{\prime}{}^0}}} \ ,
\label{nat-fins-met}
\ee
\[
         g_{ab} = {1 \over 2}
{{\partial^2 F^2} \over {\partial x^{\prime}{}^a \partial
x^{\prime}{}^b}} =  a_{ac} a_{bd} x^{\prime}{}^c x^{\prime}{}^d 
                   {\left( x^{\prime}{}^0 \right)}^{-2} \! +\,\,  
                   a_{ab} \left( \Tcal - \Ucal \right) \ .
\]

\section*{Instability and Geodesic Deviation Equation}
Within the framework of the \GDA, the tool used to investigate
the stability of the geodesic flow is given by the Jacobi--Levi-Civita ({\sf JLC})
equations for the geodesic spread, \cite{Synge}, see also 
\cite{Marco93,CPTD,MTTD}, which involve
the curvature properties of the manifold through the generalization of the
{\sl "Riemann"} curvature tensor, $R{^\alpha}_{\beta\gamma\delta}$. 
We refer to the bibliography cited above, or
also to \cite{Aquila1}, for more details. Briefly, if we want to follow
the evolution of two initially close geodesic, whose separation at the
{\sl time} $s=0$ was described by the vector of the tangent space 
$z^\alpha,\ \alpha=(0,1,\ldots,N)$ we have to find the solutions of
the \jlc equations, which, in local coordinates, read
\be 
{\displaystyle {{\nabla}\over{ds}}\left({{\nabla z^\alpha}\over {ds}}\right)}
= -R^\alpha{}_{\beta\gamma\delta} {dx^\beta\over ds} z^\gamma {dx^\delta\over ds}
= - \Hcal^\alpha{}_\gamma z^\gamma\ ,\qquad\alpha=(0,1,\ldots,N)
\ ,\label{EDG}  
\ee 
where $ \nabla/ds $
is the total (covariant) derivative along a geodesic,
\be
{\displaystyle {{\nabla v^\alpha}\over {ds}}} \defin
{\displaystyle {{dv^\alpha}\over {ds}}} + \Gamma^\alpha_{\beta\gamma} 
x^{\prime\beta} v^\gamma \ \ ;\label{dercov}
\ee
and we introduced the so-called {\sl stability tensor} $\bfHcal$, 
\cite{CPTD,Aquila1}.\\
If we consider the most general case, we have an $(N+1)$-dimensional
Finsler manifold, and the curvature tensor also depends on both coordinates
and velocity components along the geodesic, \cite{Rund}, 
\be\hskip-.6truecm
R^\alpha{}_{\beta\gamma\delta}(x^\eta, x^{\prime \eta})= 
{\left( {{\partial\Gamma^{*\alpha}_{\beta\gamma}}\over{\partial x^\delta}
}- {{\partial\Gamma^{*\alpha}_{\beta\gamma}}\over
{\partial x^{\prime \eta}}}{{\partial G^\eta}\over{\partial x^{\prime \delta}}}
\right)} - {\left( {{\partial\Gamma^{*\alpha}_{\beta\delta}}\over
{\partial x^\gamma}} - {{\partial\Gamma^{*\alpha}_{\beta\delta}}\over
{\partial x^{\prime \eta}}}{{\partial G^\eta}\over{\partial x^{\prime \gamma}}}
\right)} + \Gamma^{*\alpha}_{\eta\delta}\Gamma^{*\eta}_{\beta\gamma} - 
\Gamma^{*\alpha}_{\eta\gamma}\Gamma^{*\eta}_{\beta\delta}\ ,\label{curvtens}
\ee
where
\be
G^\alpha = {{1}\over{2}} \Gamma^\alpha_{\beta\delta} x^{\prime \beta} 
x^{\prime \delta} \ ,
\ee
\be
\Gamma^{*\gamma}_{\delta\beta} = \Gamma^\gamma_{\delta\beta}-
g^{\alpha\gamma}{\left(C_{\beta\alpha\eta} {{\partial G^\eta}\over
{\partial x^{\prime \delta}}}+
C_{\delta\alpha\eta} {{\partial G^\eta}\over
{\partial x^{\prime \beta}}}-C_{\delta\beta\eta} 
{ {\partial G^\eta}\over{\partial x^{\prime \alpha}}}
\right)}
\ee
and
\be
C_{\alpha\beta\delta} = {{1}\over{2}}{{\partial g_{\alpha\beta}}\over
{\partial x^{\prime \delta}}}\ .
\ee
If we restrict to the Riemannian case 
${\sf g}={\sf g}(x^\alpha)$, the symbols $C_{\alpha\beta\gamma}$
identically vanish, so that the {\sl Finsler connections} 
$\Gamma^{*\alpha}_{\beta\delta}$ 
reduce to the usual Christoffel symbols $\Gamma^{\alpha}_{\beta\delta}$ and
the {\sl generalized} curvature tensor $R^\alpha{}_{\beta\gamma\delta}$
becomes the Riemann tensor.
If, moreover, we consider a conservative system, there is no need, using
the Jacobi metric, to introduce the additional parameter, and all the 
indices take only the values from $1$ to $N$ (and will be used latin indices).\\

As we are interested in the global asymptotic behaviour of closeby
geodesics and we don't worry about the detailed evolution of all the components
of the perturbation, in analogy with what has been done within the hamiltonian
framework, we define an {\sl instability exponent}, which is a quantitative
indicator able to detect the possible exponential growth, in terms of
the geodesic $s$-time, of the {\sl magnitude} of the perturbation to 
the given geodesic: 
\be
\delta_I = \lim_{s \rightarrow \infty}\ \lim_{z(0) \rightarrow 0}
\left[
{\displaystyle 
{ {1\over s}\ \ln {{z(s)\over z(0)}} } 
}
\right]  \ .
\label{exp_inst}
\ee
The magnitude of the perturbation $z(s)$ is now defined in terms of 
the {\sl natural norm} on the manifold
\be
z^2(s) \equiv \|\z(s)\|^2 \defin g_{\alpha\nu}\, z^\alpha(s) z^\nu(s)\ ,
\ee
where we have understood the dependence of the metric on $s$, through its
arguments, $x^\alpha=x^\alpha(s)$ and $x^{\prime\alpha}=x^{\prime\alpha}(s)$.\\
As we have seen above, the equations of
motion of the newtonian mechanics are completely equivalent to the
geodesic equations, which give rise exactly to the same trajectories, once
rephrased the {\sl new} time $s$ in terms of the {\sl old} one $t$, exploiting
the defining realtionship amongst them, $ds=f\,dt$. 
Then we are left with a question: the equations which determine the evolution
of the perturbations within the two framework, namely the tangent dynamics
equations and the \jlc ones, are also completely equivalent each other?
Stated otherwise, the issue of the stability of motion will receive the same
answers, irrespective of the tools used to investigate it?
The answer to this question is in general negative; \ie, while the equations
which describe a single trajectory coincide, once rephrased in terms of the
same evolutionary parameter, the {\sl variational equations} do not!\\
Doubtless, in all those cases in which the issue about
the occurrence of Chaos received an unambiguous answer, we expect
that the differences among the distinct approaches do not reflect in 
qualitative disagreement. But, as remarked in the Introduction 
(see also \cite{Aquila1}), we claim that the search for the deepest insights 
on the very origin of the onset of Chaos must be performed just at the border
between  regular and unstable behaviours. From there the most intringuing
questions can arise, there the most paradoxical, and even contradictory, 
situations can occur and in this intermingled layer between Order and
Chaos we have to look in order to improve our understanding of the 
phenomena, and even, perhaps, the very definitions we are trying to
fit in the larger and larger spectrum of behaviours we are observing.
So, we expect actually the same answers from all the approaches when
the \DS\footnote{Or the region of its phase space we are investigating.} 
we are studying is either definitely regular or surely chaotic. 
Nevertheless, we cannot exclude that in the intermediate cases, 
either for a system whose global parameters assume values still in a region
in which the instability is not fully developed, or because we are exploring 
a region where the structure of the phase space is densely covered by
intermingled subsets of regular and chaotic character, the various
tools can lead to different answers.
This is exactly what happens in one single case presented below, whose 
explanation is rich of very interesting hints on the sources of Chaos in 
few dimensional \DS's and is discussed in details in \cite{CPMT_HH}.\\
Before to present the results of the comparison between the approaches,
a short list of remarks seems to be advisable.
\begin{itemize}
\item Among the undeniable merits the proposed approach possess, 
one can be fully appreciated mainly within {\sl gauge} theories, 
as in the case of General Relativistic dynamical systems, which, incidentally,
do not involve only the studies on cosmological models, but also the
analysis of the motions of particles in strong gravitational fields, as
accretion disk's particles around neutron stars or black holes, the orbital
motion of tightly bound binary stars, or even the GR corrections to the 
motion of small bodies in the solar system. 
The \GDA~is {\sl invariant by construction} with respect to any arbitrary change
of coordinates {\it and time}. 
\item This property is not shared by the usual tangent dynamics equations,
which are definitely not invariant after a time rescaling. This property
can in the general case affects only quantitatively the calculations of
the \lcn's, but in some particular situations can even change the qualitative
outcomes, even when the rescaling is very smooth.
\item To see why the usually adopted equations for small perturbations 
are not the most natural ones, to which to allot some privileged role, 
it suffices to look at the more or less equivalent geometrization 
leading to the {\sl Eisenhart metric}, \cite{Eisenhart,Marco93}.
In this case, indeed, one can recover a form of the \jlc equations which
is completely equivalent to the tangent dynamics ones, through an {\sl affine
parametrization} of the geodesics. But, from the derivation itself, it
is clear that this choice is only one among many possible.
\item The Eisenhart geometrization can be obtained also from a slight
generalization of the procedure outlined above, only for $N$ degrees of
freedom conservative holonomic \DS's, starting from the
$(N+1)$-dimensional configuration space-time, and introducing a further
parameter, $s_E$, just in order  to achieve an {\sl affine relation} 
between the latter and the newtonian time, $t\sim x^0$. 
The manifold then turns out to have $(N+2)$ dimensions and the 
line element reads 
\be 
ds_E^2 = a_{ij} dx^i dx^j - 2 \Ucal({\bf x})\,
        (dx^0)^2 + 2 dx^0\, dx^{N+1} \quad ,
\label{ds2E}
\ee
where $x^{N+1}$ is the additional coordinate, that {\it can} be 
chosen as
\be
x^{N+1} (t) = A^2 t + B - \Int_0^t L\, dt\ \ ,
\label{act}
\ee
with $A$ and $B$ constant. As remarked above, {\sl with this choice}
the parametrization turns out to be affine:
\[
 ds_E^2 = A^2\,dt^2 \ .
\] 
\item Nevertheless, the possibility to frame the tangent dynamics equations,
\ie, one of the most used tools to investigate the chaotic behaviour of
\DS's, into a geometric picture, is in itself a neat indication of the
relevance the \GDA~can hold and a spur to address to it the attention 
deserved.
\item We finally emphasize the wider applicability of the Finsler
approach, which is not limited to holonomic systems, and whose
reliability will turn out clearly in the applications we sketch below.
\end{itemize}
We now present some results (for a complete account on them, see
\cite{CPMT_HH} and \cite{MTCP_BIX}, respectively) obtained from
the application of both the \GDA's and the tangent dynamics equations
to the two few dimensional dynamical systems mentioned before. In order to
have a more understandable check about the equivalence of equations, we
get rid (only here!) of the covariant form of the \jlc equations, 
exploiting the definition of {\sl total covariant derivative}, 
eq.(\ref{dercov}), and of the curvature tensor, eq.(\ref{curvtens}), 
and expressing all the derivatives in terms
of the newtonian time $t$.\\
The equations to be compared with the tangent dynamics equations, 
eqs.(\ref{tvar}), result much more involved when expressed in a form
which is not covariant, and in terms of a parameter which is not
the natural one. It is clear nevertheless that we couldn't
complete the steps in the opposite direction: while the change
of the time parameter can be performed in the backwards direction,
it not possible to start from eqs.(\ref{tvar}) and to give them 
a covariant form.\\
When the two steps have been done, and assuming for sake of simplicity
that the kinetic energy form is euclidean $(a_{ij}=\eta_{ij})$, 
the \jlc equations, (\ref{EDG}), in the Jacobi and Finsler metrics read, 
respectively,
\be
\ddot z^i +\Ucal_{,ij} \ z^j = - {\Ucal_{,i}\over E-\Ucal}
(\Ucal_{,j}z^j+{\dot x}^p{\dot z}^p)
+\dot{x^i}\left[ z^p {\partial\over \partial x^p}
\left( {1\over E-\Ucal}{d\Ucal\over dt}\right)
+ {\dot z}^p{\partial\over \partial {\dot x}^p}
\left( {1\over E-\Ucal}{d\Ucal\over dt} \right)
   \right] \ ;
\label{sjvar}
\ee 
and
\be
\ddot z^i +\Ucal_{,ij} \ z^j = - 2 \dot z^t \Ucal_{,i} + 
\ddot z^t \dot x^i\ ,
\label{sfvar}
\ee
where
\be
\ddot z^t = 4 {\dot z^t\over L}{d\Ucal\over dt} (1+\Ucal)
+2 z^i {\partial\over \partial x^i}
\left( 
{1\over L}{d\Ucal\over dt}
\right)
+ 2 \dot z^i {\partial\over \partial \dot x^i}
\left( 
{1\over L}{d\Ucal\over dt}
\right)\ .
\label{sfvart}
\ee
The equations for the disturbances in the Finsler case have been splitted
for $\alpha=(1,\ldots,N)$ and $\alpha=0$, and we remark that the
latter appear at the \rhs of equation (\ref{sfvar}), \ie, the
{\sl acceleration} of the
$0^{th}$ component of the perturbation, influences all the
other {\sl spatial} components.\\
From a direct comparison of the equations above, 
(\ref{sjvar})$\div$(\ref{sfvart}), a striking difference with respect to 
the {\sl simpler} eqs.(\ref{tvar}) is apparent.
Nevertheless, as emphasized above, we expect that they give the same results
when used to detect instability in situations in which the behaviour is
either markedly regular or strongly chaotic. And this is actually what happen. 
\section*{Numerical applications}
In this section we present some of the results obtained by numerical
integration of the dynamics, obtained either from equations of
motion (\ref{tmotion}) or from the geodesic equations, (\ref{Geo}) 
along with the simultaneous numerical integration of the
equations for the perturbations, eqs.(\ref{tvar}) and 
(\ref{sjvar})$\div$(\ref{sfvart}) for two \DS's, namely  
the H\'enon--Heiles Hamiltonian and the Bianchi IX model.
\subsection*{H\'enon Heiles dynamics}
This {\it simple} dynamical system has raised to the role of a 
{\sl testing ground} for any proposed method to investigate the
chaotic dynamics of few dimensional \DS's in general,
\cite{BGS76}, and in particular
it has become a paradigmatic example of the process of onset (or, perhaps
better, the {\sl diffusion}) of
Chaos along with the variation of a parameter of the system, \cite{HH},
in this case the energy itself playing this part.\\
As it is well known, the H\'enon-Heiles ({\sf HH}) potential represents
 a two dimensional system introduced more than
three decades ago, \cite{HH}, in order to explain some features
of the motion of stars in a galactic potential. Since then, as said
above, after the first surprising results obtained there and in subsequent
studies, its relevance as an astronomical models has been entirely
overshadowed by its intriguing features as \DS. Among its interesting
properties, because the relevance it holds in the present context,
it emerges the rather abrupt transition from a quasi completely
integrable dynamics at low energy, $E\minord 0.1$,  to a relevantly
chaotic behaviour above this threshold, $0.1\minord E\leq 0.166$, up
to the escape energy $E=1/6$. The origin of this sudden modification
of the qualitative behaviour of trajectories has been investigated
in details; amongst them we recall the Toda's attempt, \cite{Toda}, towards a
{\sl synthetic} criterion to detect the onset of instability. While
unsuccessful, \cite{criticaToda}, that work spurred a lot of efforts
devoted to an understanding more than phenomenological on the
dynamical instability, \cite{Marco_HH,CPMT_HH}.
It is also interesting to note that below the stochasticity threshold
the classical perturbation theory gives excellent agreement with
the outcomes of {\sl numerical experiments}, \cite[\S 1.4a]{Lichtenberg1},
whereas above that level the numerical phase portrait of the system
turns out more and more different from that obtained 
analitically. This sudden {\sl breaking down} of another
{\sl quasi conserved} quantity can help to shed light also on
the behaviour of higher dimensional \DS's.\\
The H\'enon-Heiles Hamiltonian reads
\be
H = {\dot x^2\over 2} + {\dot y^2\over 2} + \Ucal (x,y) \ ,
\ee
with potential
\be
\Ucal (x,y) = {{1} \over {2}} \left( x^2 + y^2 \right) 
   + x^2 y - {{1} \over {3}} y^3 \ .
\label{HHpot} 
\ee
We report in figure \ref{fig1} the well known, \eg,\cite{BGS76},
behaviour of \lcn's versus time (in a log-log scale) at the intermediate
energy $E=0.125$, for which there is a {\sl coexistence} between 
regular and chaotic zones, for a representative set of initial conditions, (i.c).
 As we can see, the signature of Chaos is neatly detected, looking at the
markedly different behaviours of the \lcn's. After some transients, 
and we remember that \lcn's are indeed asymptotic indicators, 
the definitely regular orbits, labeled by the first two digits, 1 and 2,
 show an unambiguous behaviour indicating the absence of any 
exponential instability:
\be
LCN(t) \sim  {{\ln{t}}\over{t}} \vainf{t} t^{-1}\ ,\qquad 
{\rm for\ i.c.\ 1\ and\ 2\ .}
\ee 
Analogously, the three {\sl surely} chaotic orbits, emanating from i.c.
4$\div$6, show a clear-cut indication of a common exponential
rate of growth of the {\sl euclidean norm} of the deviation vector $\D$.\\
At variance, the initial condition labeled as 3 manifests a behaviour which is
much more noisy than the other orbits, though it seems to give a nevertheless
convincing indication of stable behaviour. We tested the peculiarity of this
and associated orbits, \ie, those emanating from the same island on the 
Poincar\'e surface of section, $x=0$, ({\sf PSS}), looking at all the possible 
sources of the fluctuations, and even implementing a version of the 
{\sl renormalization procedure} slightly modified with respect to the 
{\sl universally} adopted one, proposed by \cite{BGS76}, which occasionally 
proved to be more convincing, see \cite{CPMT_HH}. 
Moreover, while all the other initial
conditions tested do not show any perceptible dependence with respect
to a change on (the orientation of) the initial perturbation, $\D(0)$,
the behaviour of the \lcn~for orbit 3 is sensitive, in the absolute 
value as well in the overall behaviour, to this choice, as shown by the curve 
labeled as ${\rm 3}^{\prime}$ in fig.\ref{fig1}, obtained starting from the 
same i.c., but selecting a {\sl very bad} orientation of the initial 
perturbation. With this choice the noise is quite relevant, neither is 
sensibly reduced performing a longer integration run, as can be seen looking
at the small panel in the  same figure.\\ 
On the basis of these outcomes, we claim that, though almost certainly
{\sl regular} according to the accepted {\sl canonical definitions}, the orbits
emanating from the island(s) on the $\dot y$ axis in the {\sf PSS}, possess some
characteristics which make them qualitatively slightly distinct with respect 
to the ones belonging, for example, to the main island on the $y$ positive
semi-axis or also to the orbits starting from inside the 
{\sl banana shaped} region on the $y<0$ side. 
As a matter of fact, we can roughly guess from figure \ref{fig1} that 
the behaviour of the perturbation for this set of i.c., is not simply
polynomial in time, but shows some intermittent phases of instability,
probably of {\sl stretched exponential} nature.\\
This initially regarded as {\sl annoying} behaviour has been instead revealed
very fruitful, as it has persuaded to look more carefully to the
proposed {\sl geometrical origin} of instability 
and allowed us to find some convincing explanations for those 
peculiarities, which turn out to be very enlightening for a deeper
understanding of the sources of Chaos, \cite{CPMT_HH}.\\
As heralded above, we do not expect to find any relevant difference in
analyzing the qualitative properties of the dynamics of the H\'enon
and Heiles system, when it displays a well defined nature.
And actually we can see from figure \ref{fig2} how the {\sl instability
exponents} defined in eq.(\ref{exp_inst}) for the Jacobi metric give,
in such situations, substantially the same answers.
In fig.\ref{fig2} we have plotted the instability exponents,
$\delta_{I_J}$, (indicated simply as $\delta_J$ in the figure) obtained by the
numerical integrations of the geodesic equations for this
metric, along with the corresponding geodesic deviation equations, in terms
of the parameter $s_J$ defined through the relation
$ds_J = \sqrt{2}\Tcal\,dt$, in turn integrated 
simultaneously to the previous ones, in order to stop the run at the same
value of the newtonian time $t$ reached with the tangent dynamics equations.
A part from a trivial numerical multiplicative factor, related to the non
affine reparametrization, because of which the Jacobi parameter is always
smaller (for the {\sf HH} system at this energy) than the newtonian time, 
and consequently the instability exponent
turns out to be greater than the \lcn~by the inverse factor\footnote{In the 
following formula the estimate on the average value of the kinetic
energy  is obtained exploiting the {\sl virial theorem} in an approximated 
form.},
\[
s_J(t) = \sqrt{2}\,\Int_0^t \Tcal (t') dt' = \sqrt{2} \langle\Tcal\rangle_t\, t
\cong {E\over{\sqrt{2}}}\, t 
{\raisebox{-1.1ex}{$\stackrel {\displaystyle {\llongrightarrow } }
{\scriptscriptstyle{E = 0.125 }}$}}\ 0.089\,t \sim t/10\ ;  
\]
we find out a completely satisfactory agreement for the {\sl truly chaotic}
orbits, whereas, for the {\sl quietly regular} ones, the qualitative
agreement is equally convincing, though the behaviour of the $\delta_{I_J}$
is more noisy that that of the corresponding \lcn's. 
But the most striking difference concerns the answer given by the Jacobi
\GDA~to the issue of the regularity of orbits of the (upper) island in the
{\sf PSS}: according to fig.\ref{fig2}, orbit with i.c. labeled as 3 turns
out to possess a strong dependence on initial conditions!
In order to discard the possibility of an artifact due to a too short
integration time, in the small frame in figure \ref{fig2} we plot the long
term behaviour of the $\delta_{I_J}$, which highlights once more the 
firm decision of Jacobi geodesic deviation equations to depict as chaotic
that orbit.\\
While it is not difficult to understand to source of noise in the $\delta_{I_J}$
of regular orbits, 1 and 2, it turns out a definitely not easy task to single out
the sources of such a surprising behaviour of the third one. The answers
to these questions reside partially on the singular behaviour of the conformal 
factor $f_J(\x,\dot\x)$ and consequently of the Jacobi metric itself, 
when the kinetic energy vanishes.
 Such an occurrence can be ruled out in the case of many degrees of freedom 
systems, \cite{Aquila1}, but cannot be excluded in the case of few dimensional 
\DS's, as in some cases the orbit can touch the {\sl zero velocity surface} 
(or {\sl curve}, when $N=2$), which is the boundary, for {\sl standard} 
\DS's\footnote{As we will see below,
this is not the case for General Relativistic \DS's, for which the 
{\sl zero kinetic energy surface} do not coincide with the surface of
{\sl zero velocities}, as the kinetic energy form is not positive definite. Moreover,
for standard (conservative) \DS's, the surface of zero kinetic energy is the
$(\Ncal-1)$-dimensional manifold $\{\q,\p\} : \{\Ucal(\q)=E\, ;\, \p={\bf 0} \}$,
whereas, for non positive definite kinetic energy forms, the $\Tcal=0$ manifold
has in general dimension $(2\Ncal-2)$: $\{\q,\p\} : \{\Ucal(\q)=E\, ;\, 
a_{ij}p^i p^j= 0 \}$.\label{nota_KEnonpos}}, of the region allowed to motion.\\
These singularities suffice to explain the oscillatory behaviour of instability
exponents of the two regular orbits, but alone cannot help to grasp the
origin of such a {\sl spiteful} performance of the third one.
Actually its very peculiar behaviour has been for us a very strong push to go
beyond the first attempts to relate curvature and stability, and provided
very deep hints on the relationhips between geometry and chaos, about which
a detailed account can be found in \cite{CPMT_HH}.
Incidentally, we note that in the very interesting paper by \cite{Marco_HH},
where the {\sl qualitative} character of orbits has been investigated by means
of an equation derived by the \jlc ones, the behaviour of this (or similar) orbit 
has not been analyzed, whereas, here, we recover (see below) also a complete 
{\sl quantitative} agreement (once exploited the relationship
among $s_J$ or $s_F$ and $t$) between the tangent dynamics equations and the 
{\sl exact} \jlc equations,  in the entire {\sf PSS} $x=0$. In
\cite{CPMT_HH}, we moreover discover a {\sl geometric indicator of instability} 
able to go beyond the Toda criterion, in that it can distinguish among chaotic and
regular orbits, {\sl at the same energy}.\\
The singularity of the Jacobi metric mentioned above do not occur within
the  Finsler {\sf GDA}, as the relationship between the geodesic parameter
and the newtonian time is governed by the function 
$F(x^\alpha,x^{\prime\alpha})$ defined above, which is derived
from the lagrangian of the system, $L(\x,\dot\x,t)$. 
It is well known, \eg\ \cite{Goldstein},
that, given a \DS, the lagrangian function is defined up to an additive
{\sl gauge function}, $m(\x,t)$, which must only fulfil the condition to be
the total derivative with respect to time of any differentiable function 
$M(\x,t)$ defined on configuration space.\\
In the numerical study of the geodesic flow derived from the {\sf HH}
potential, we actually exploited this freedom in order to obtain 
an {\sl average} equality among newtonian time and Finsler parameter; \ie,
we chosen $\bar{L}(\x,\dot\x,t) = L(\x,\dot\x,t) + m(\x,t)$ and consequently
\[
s_F (t) = \Int_0^t \bar{L}[\x(t'),\dot\x(t'),t']\, dt'\, =\,
          \Int_0^t      L [\x(t'),\dot\x(t'),t']\, dt'\, + M[\x(t),t] -
          M[\x(0),0]
\]
\noindent in such a way that $\langle\bar{L}\rangle_t \cong 1$, \ie, on
the long run, $s_F(t)\cong t$. This choice obviously didn't modified the
results, but only allowed a direct quantitative comparison between the
{\sl digits} of the Finsler instability exponents and the \lcn's.\\
In figure \ref{fig3} we show the $s_F$-behaviour of the $\delta_{I_F}\!$'s 
(they also simply indicated as $\delta_F$ in the figure), for the same 
set of i.c. used to compute the \lcn's and the $\delta_J$'s.
The complete, quantitatively detailed, agreement with the results of the 
tangent dynamics equations is strikingly evident for all the i.c. analyzed,
except again for the orbit with label 3. 
But, at variance with what occurred in the Jacobi description, now the
results achieved within the \GDA~show a much less noise than the \lcn's! And
the answer about the issue of instability of these initial conditions is 
strikingly definite. Moreover, we tested the behaviour of the $\delta_F$'s 
varying the initial
orientation of the perturbation, being not able to distinguish between
various choices, already at relatively small values of $s_F$.\\
Nevertheless, we didn't content ourselves with the nice performances of the
Finsler \GDA~and investigated both the reasons for a so satisfactory
result and the causes at the origin of the seemingly deceptive outcomes
of the Jacobi description. As remarked above, this at first sight misleading
result of the Jacobi approach, has provided very useful insights on the
analysis on the sources of instability of \DS's and of its relationships with
the curvature properties of the manifold. A full account of the results
achieved is presented in \cite{CPMT_HH}.

\subsection*{Bianchi IX dynamics}
In order to point out once more the power of the \GDA~and, at the
same time, the cautions needed in order to avoid a too literal
extrapolation to realistic \DS's of the results borrowed
from the abstract Ergodic Theory, \eg,\cite{Chitre,Szydlowsky}, we
then performed a numerical study of the dynamics of a General
Relativistic three degrees of freedom dynamical system, which
share with the H\'enon-Heiles hamiltonian the strange destiny
of becoming ever and ever more interesting as a paradigmatic 
\DS~within the theoretical context to which belongs 
(Celestial Mechanics and General Relativity, respectively), than
as realistic model of the physical phenomenon it was intended
to describe originally.\\
The interest of this \DS~relies on the lively debate it raised among
the cosmologists, since the last three decades, \eg,\cite{Bel}, 
and more recently even between people working on Ergodic and 
Dynamical Systems Theories, \cite{Khalatnikov}
and \cite{Contopoulos,Hobill_book}. 
Indeed, amongst the mess of studies performed until recently, 
of both analytical and numerical character, about half of them
were reporting {\sl convincing evidences} on the regular
 features of the {\sf BIX} dynamics; whereas the 
others claimed the existence of {\sl strong indications} about the
chaotic properties of the model. In this very last period, it
seems to be in fashion the belief that this \DS~is indeed chaotic,
\cite{Cornish}, also because some amongst the previous results indicating 
{\sl perhaps} even 
its integrability have been corrected, \cite{Contopoulos}.
 It had been already suggested that these disagreements depend
on the different time parameter adopted to describe the dynamics,
\cite{Hobill}. 
Actually, in a General Relativistic context, the {\sl gauge freedom}
 in the selection of time and coordinates can be exploited with 
much more liberties, as the results should be {\sl invariant} 
with respect to the choice made.\\
The dynamics of the so-called Bianchi IX cosmological model\footnote{
See, \eg,\cite[\S\S 115$\div$118]{LL2} and \cite{Hobill_book} and 
references therein for a description of its original 
motivations and a discussion on the issues still open.}, 
can be studied with a variety of approaches, although the most suitable
from a \DS~viewpoint is actually the hamiltonian one.\\
With an appropriate choice of variables, the \BIX~Hamiltonian reads
\be
H =  {1\over 2}\dot\beta_{+}^2 + {1\over 2} \dot\beta_{-}^2 
- {1\over 2} \dot\alpha^2 +
     \Ucal(\alpha,\beta_+,\beta_-) \ ,\label{primaH=0}
\ee
with potential
\be
 \Ucal ={1\over 8}e^{4\alpha} \left\{ {1\over 3}e^{-8\beta_+} - {4\over 3}
e^{-2\beta_+} 
   \cosh\left(2\sqrt{3}\beta_-\right) + {2\over 3}e^{4\beta_+}
   \left[ \cosh\left(4\sqrt{3}\beta_-\right) - 1\right]\right\} \ ,
\label{pot_BIX}
\ee
where $\beta_+$, $\beta_-$ and $\alpha$ are functions of the scale factors 
$a$,$b$,$c$ of the Universe, and measure the overall volume and the 
anisotropy of the $3$-dimensional hypersurface $\tau=const.$;
 the dot indicates differentiation with respect to
the time parameter $\tau$, defined in terms of the {\sl cosmological proper time}
$t$ through the transformation $d\tau=dt/abc$.
The Hamiltonian (\ref{primaH=0}) in nevertheless peculiar, in that the
General Relativity imposes a constraint not encountered usually in classical
\DS's:  Einstein equations request that $H$ be a {\sl null hamiltonian}, \ie,
$H\equiv 0$.\\
The possibly chaotic properties of this relativistic dynamical
system could be a signal about the occurrence of some kind of
{\sl statistical} behaviour of the primordial Universe, able to
explain some observational evidences without the introduction
of {\sl exotic} theories, as the Inflation. 
Nevertheless, as remarked above, in the last years, it has assumed 
importance as a dynamical system in itself, because of the
strikingly conflicting answers about its
chaotic behaviour.
It has become clear that the discrepancies could arise from 
different choices of the time parameter, and even some analytical
results obtained by a {\sl discretization} of the flow through 
the introduction of different maps, can hide some inaccurate
approximantions, which turns out to be responsible of
the misleading conclusions about the stochastic character of the
dynamics. We refer to \cite{Hobill_book} for an exhaustive review
of the existing treatments, and to \cite{Sigrav96} and \cite{MTCP_BIX},
respectively, for preliminary and detailed results.\\
General Relativity requires covariance and so
it is essential to analyze the Bianchi IX dynamics with methods
whose results are independent of the choice of the coordinates, and, in
particular, of the time variable. For this reason, we decided
to implement the \GDA~to the study of its possibly chaotic
behaviour. At the point we should make a comment: as it is well
known, in order to properly speak about {\sl Chaos}, it is
necessary that the dynamics possess a strong dependence on
intial conditions and moreover that the ambient space where
the dynamics is decribed (the phase space within the Hamiltonian
treatment) must be {\sl compact}, \ie, the {\sl stretching} caused
by the exponential instability, must be accompanied by the
{\sl folding} due to the bounds imposed to the distance between
any two point in this space by {\sl compactness}. Looking at
the Hamiltonian (\ref{primaH=0}), we see at once
that the second condition mentioned above is not fulfilled by
the \BIX~dynamical system; so, even if we could be able to prove the
occurrence of an exponential instability, it should not be rigorous to
speak about Chaos, though it can be argued  that when the dynamics shows
an exponential instability characterized by a given time-scale $\tau_i$, 
if the trajectory of the system remains {\sl trapped} in an unstable region
of phase space (or something else) for an interval of time $T_{tr}$ much longer
than the instability time, $T_{tr}\gg\tau_i$, then the effects on the
predictability of the evolution would result substantially the same
as if it were a {\sl truly} chaotic dynamics. We will see below, however, that
in this case we don't need to deepen the reliability of this argumentation,
as the \BIX~dynamics do not shows any exponential instability!\\
The second preliminary observation concerns the possible geometrizations
of this \DS. It was already recognized that a geometrical description of
the \BIX~dynamics could help to get rid of any {\sl gauge dependent} 
approach, \cite{Chitre,Szydlowsky}. Nevertheless, these authors didn't
paid enough attention to the peculiar features imposed by the General
Relativity to this particular \DS.
Indeed, the Hamiltonian (\ref{primaH=0}), though in its own way, describes
however a conservative dynamical system. On this light,
in order to give a geometrical description of the dynamics, those authors
exploited the Jacobi metric, using the line element
defined above, \ie, the invariant Jacobi parameter
\be
ds_J= \sqrt{E-\Ucal}\, \sqrt{2\Tcal}\, d\tau \ ,
\ee
where $\Ucal$ is the potential defined in (\ref{pot_BIX}) and the
{\sl kinetic energy} $\Tcal$ turns out from (\ref{primaH=0}):
\be
T = {1\over 2}\dot\beta_{+}^2 + {1\over 2} \dot\beta_{-}^2 
- {1\over 2} \dot\alpha^2 \ .
\ee
It is evident why this treatment cannot be pursued further: it is true
that the Jacobi parameter is invariant, but this make even more
serious its singularities, repeatedly occurring along the evolution
of the system, when the trajectory cross the {\sl zero kinetic energy}
surface ${\cal S}_{\Tcal=0}$, which, because of the Hamiltonian constraint,
coincides also with the surface where the potential $\Ucal$ vanishes.
As heralded above, this \DS~gives an example of non coincidence between
the {\sl zero kinetic energy} and {\sl zero velocities} surfaces. Now,
actually, the former no longer constitutes the {\sl external} boundary
of the region allowed to motion in configuration space, but it is
instead inside that region, so it can be crossed by the actual trajectory
infinitely many times during the evolution, and with whatever incidence
angle. The comments made in the footnote \ref{nota_KEnonpos} about the different
dimensionalities of the manifolds of vanishing kinetic energy for standard
and pecular \DS's help to intuitively grasp the very different chances of such an
occurrence in the two situations. Incidentally, we observe that the interpretation 
of the results obtained within this (though unreliable) approach are nevertheless 
easily understandable: as the kinetic and potential energies along a trajectory
equal each other, except for the sign, and as the potential is appreciably
different from zero only near the {\sl walls}, \cite{MTTD,PRE97a},
we see at once that the Jacobi description amounts essentially to
reduce the continuous flow to a discrete bouncing map, neglecting the
ever increasing amount of time elapsed between successive bounces. This
leads to an overestimate of the instability rate, which is discussed in
details in \cite{MTCP_BIX}.\\
So, while the Jacobi \GDA~gives very satisfactory performances when 
applied to standard \DS's with many degrees
of freedom (see also \cite{Aquila1}), it breaks down when applied to 
{\sl peculiar} few dimensional spaces.\
These difficulties of the geometrical method have raised some
criticisms against the overall approach, \eg,\cite{Burd}, and
discouraged for some time any further attempt.
But, in these cases again, the \GDA~on Finsler spaces reveals all 
its potential, allowing a geometrical, invariant by construction, 
transcription of the dynamics even of strongly singular \DS's.\\
Indeed, in figure \ref{fig4} we plot the instability exponents 
$\delta_{I_F}$, computed using the Finsler metric, for a small number
of typical orbits, chosen from a nevertheless representative set of
initial conditions, leading to distinct overall behaviours (for details
see \cite{MTCP_BIX}).
It is evident  at once that we can discard the possibility of the occurrence 
of any global exponential instability.
The presence of some {\sl light peacks}, which become ever and ever 
smaller going towards the singularity\footnote{We recall that the \BIX~model 
represents a {\sl closed universe}, with a {\it Big Bang}, followed by an
anisotropic expansion which becomes more and more isotropic towards the maximum,
and then a collapse to a {\it Big Crunch}, which tends to anisotropize again 
for $|\tau|\rightarrow\infty$.}, can be detected looking carefully at the
$s_F$ behaviour of the instability exponents. These positive peacks, which tend 
for a while to stop the almost uniform decrease of $\delta_{I_F}$,
occurs every time there is a transition from one
{\it Kasner epoch} to another one, or within the {\sl minisuperspace} picture,
\cite{Hobill_book}, to the bounces of the trajectory against the
potential walls. They reflect the occasional occurrence of an instability
phase whose duration is however very small with respect to the ever
increasing interval between two successive phases. On this light, 
this asymptotic behaviour of the {\sl instability exponents} could 
partially support some recent claims
about the occurrence in the \BIX~dynamics of a sort of 
{\sl chaotic scattering}, as suggested in \cite{Contopoulos} and \cite{Cornish}.\\
For a more detailed analysis of this and other issues, we refer to
 \cite{MTCP_BIX}, where it is also
investigated the relationships existing between the dynamical
features of th \BIX~model and the curvature properties of the
associated Finsler manifold.

\section*{Conclusions}
In this paper we presented an alternative derivation of the
procedure which leads to the transcription of the dynamical properties
of general lagrangian systems in geometrical terms.\\
Using this {\sl elementary} derivation, we shown the complete
equivalence of first order variational principles, leading to the
equations of motion and to geodesic equations, within the two 
frameworks, respectively. Then we addressed to the issue of
the possible equivalence of second order variational equations, which
convey the informations related to the equations describing the behaviour 
of the {\sl disturbances}, \ie, the perturbation vectors in phase space,
and the geodesic deviation vectors on the tangent space of
the manifold, whose evolution is determined, respectively, by
the {\sl usual} tangent dynamics and the Jacobi--Levi-Civita
equations. We compared the explicit expression of the former 
with those resulting by two specific geometrical transcriptions, and,
once recognized the non coincidence, we tested their however
possible equivalence through the application to two different
but paradigmatic \DS's.\ We found a completely satisfactory
agreement between the results obtained in the usual hamiltonian setting
and those coming from the \GDA. Moreover, the geometrical tools 
for the investigation of the stability properties of the dynamics
show nevertheless some minor differences which, when fully exploited,
bring very enlightening suggestions on the relationships between
the curvature properties of the manifold and the possible onset of
Chaos.\\
A full account of the results of the {\sl numerical experiments} along with
the implications of these investigations will be presented 
in \cite{CPMT_HH}, devoted to the
H\'enon-Heiles Hamiltonian, and in \cite{MTCP_BIX}, for 
the Bianchi IX dynamical system. They, along with the existing results, 
obtained by other researchers, on both many degrees of freedom systems, 
\cite{Marco93}, and few dimensional ones, \cite{Marco_HH}, and
by the authors, \cite{CPTD,MTTD} and \cite{PRE97a,Aquila1}, strongly
support the claims on the reliability of the \GDA, and nevertheless
point out the need of a careful check of the conditions needed in order
to apply safely the geometrical transcription of the dynamics, to fully
exploit the power of a tool which can give very deep insights
on the origin of instability in \DS's, and also interesting hints
to single out better criteria and/or indicators of chaotic behaviour,
in particular in all those situations, and there are many of such a kind,
where the boundary between Order and Chaos is so nuanced. 

\begin{center}
{\Large {\bf FIGURE CAPTIONS}}
\end{center}

\begin{figure}[h]
\caption{LCN's for the H\'enon--Heiles Hamiltonian 
at $E=0.125$.\ The initial conditions corresponding to each labeled
curve are indicated in the figure. The curves labeled as
3 and ${\rm 3}^{\prime}$ refer to the behaviour of the LCN computed along
the same orbit but with two different choices of the initial perturbation. 
The small frame shows the behaviour of these curves for an interval of time
ten times longer. In this panel, the curve labeled as 3 has been
shifted upwards, to avoid ovelapping.\label{fig1}}\vskip1cm

\caption{Instability exponents $\delta_{J}$, calculated
in the Jacobi geometry for the HH system. The initial 
conditions are chosen and labeled as in figure \ref{fig1}. 
The differences in the vertical  and horizontal scales with respect 
to those of figure \ref{fig1} are simply due to the relation between 
$s_J$ and $t$ (see text). 
The inserted frame shows the long time behaviour (up to $t=10^5$) of 
the instability exponent of the orbit with initial conditions labeled as 3; 
note that the vertical scale has been rescaled
in order to highlight the smaller and smaller fluctuations\ .
\label{fig2}}\vskip1cm

\caption{Instability exponents $\delta_{F}$, calculated
in the framework of Finsler gemetrodynamics. 
The initial conditions are chosen and labeled as in figure \ref{fig1}\ .
\label{fig3}}\vskip1cm

\caption{Short (upper panel) and long (lower panel) time behaviours
of the instability exponents, $\delta_{I_{F}}$, computed in
the Finsler framework for the Bianchi IX dynamical system.
The initial conditions corresponding to each curve have been
chosen to be representative of the overall set of
{\sl typical} behaviours of this system.
It is evident the signature of a {\sl regular character} of the
dynamics, and also the {\sl stretching} of time intervals between
two successive peaks\ .
\label{fig4}}
\end{figure}

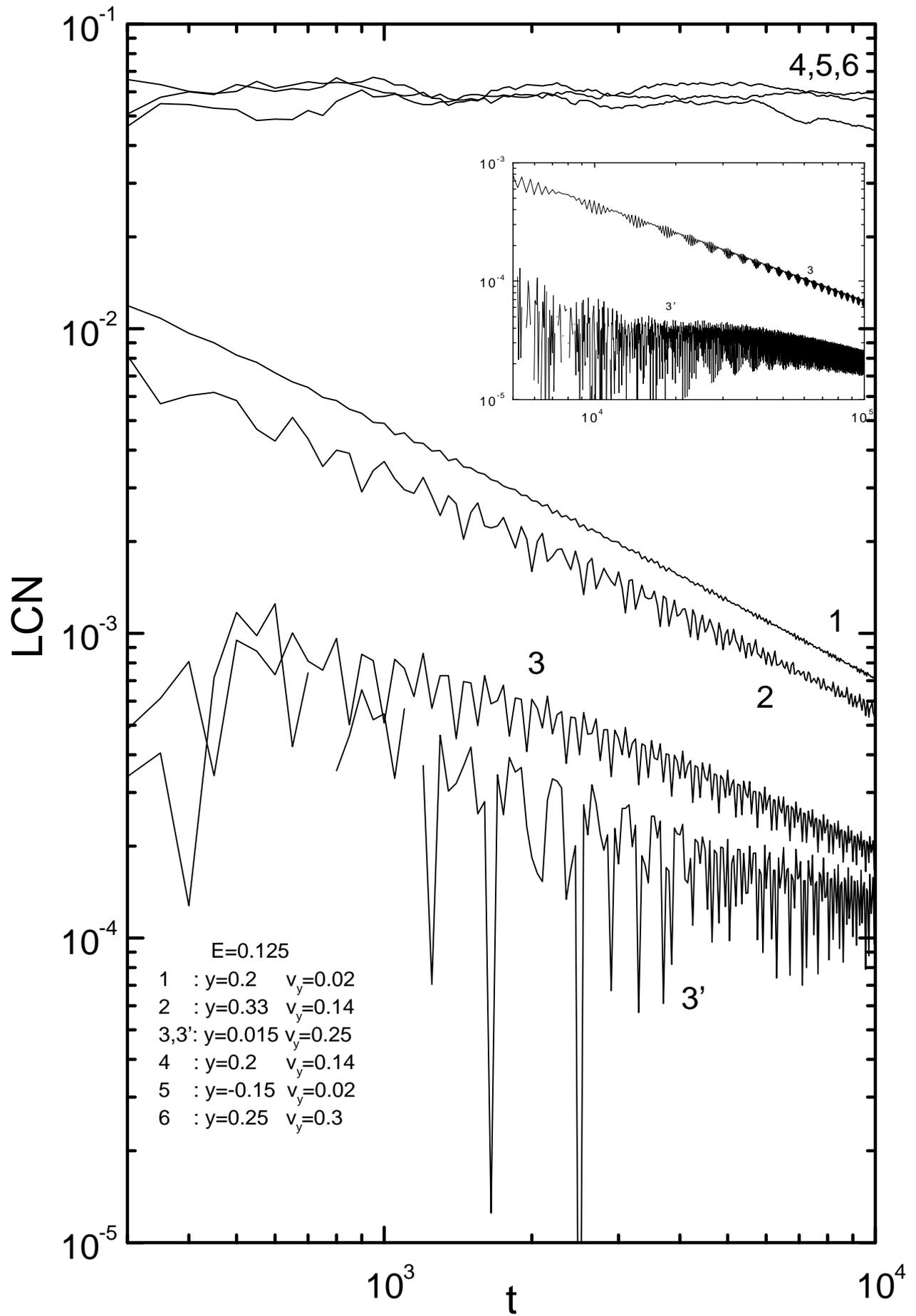
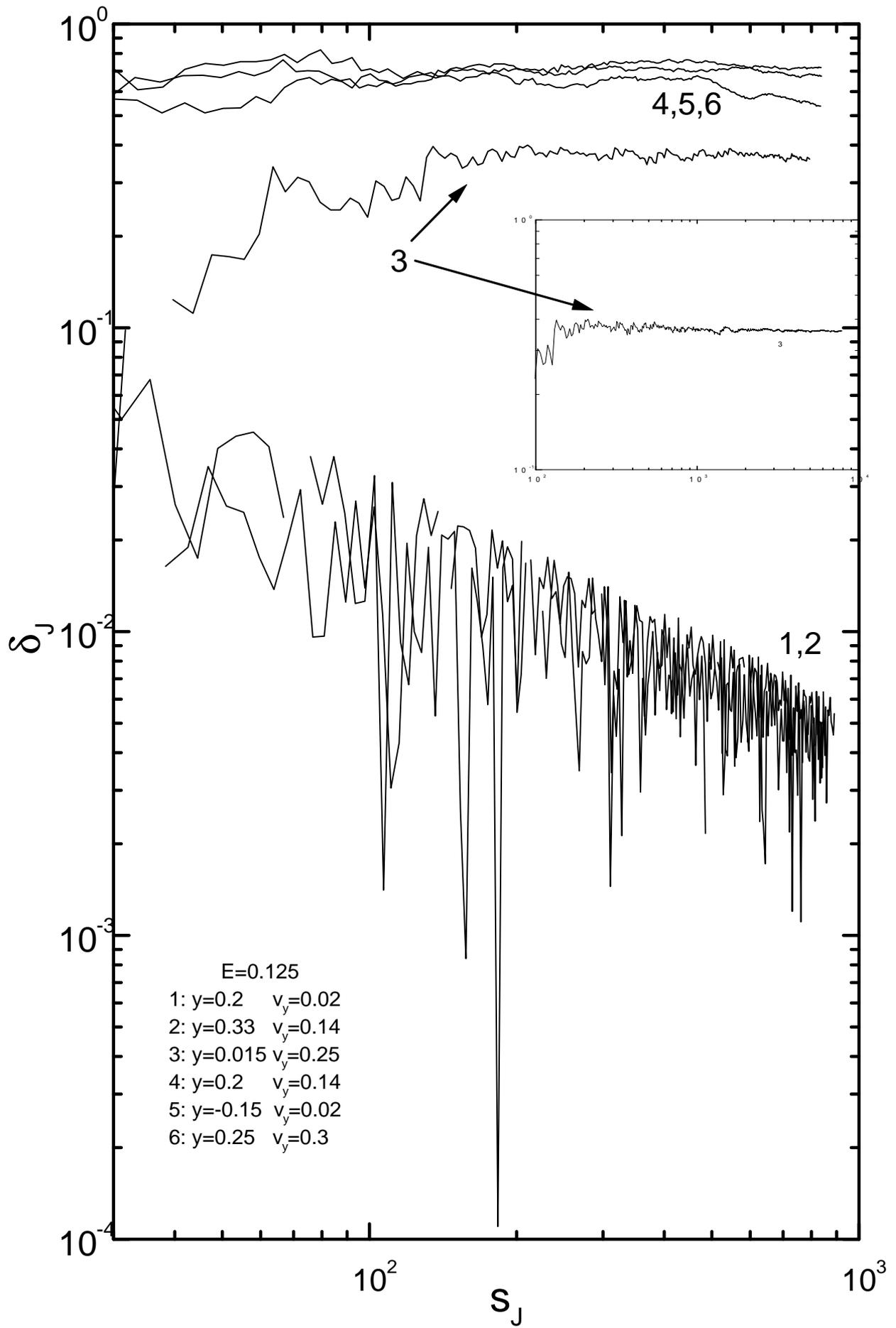
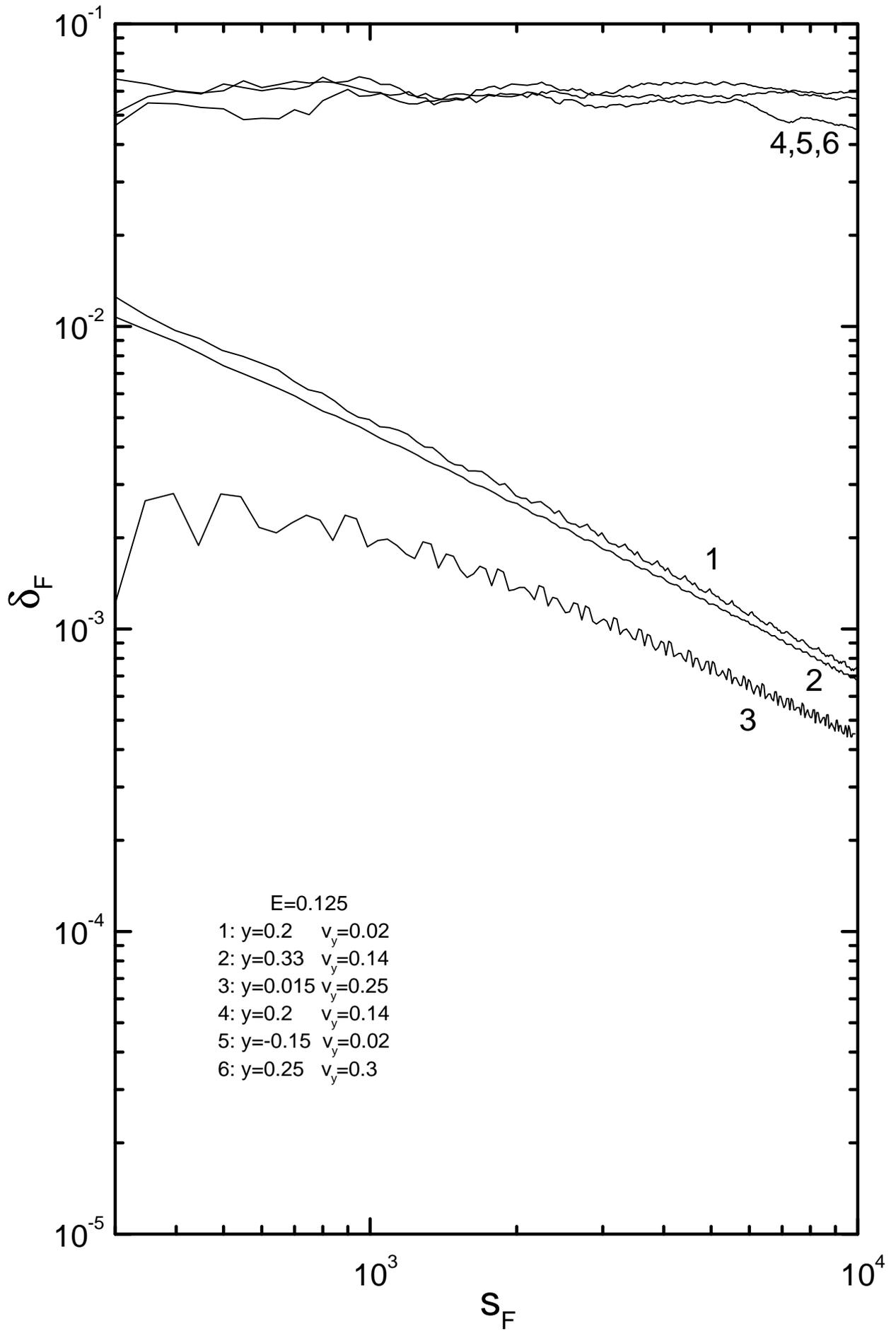
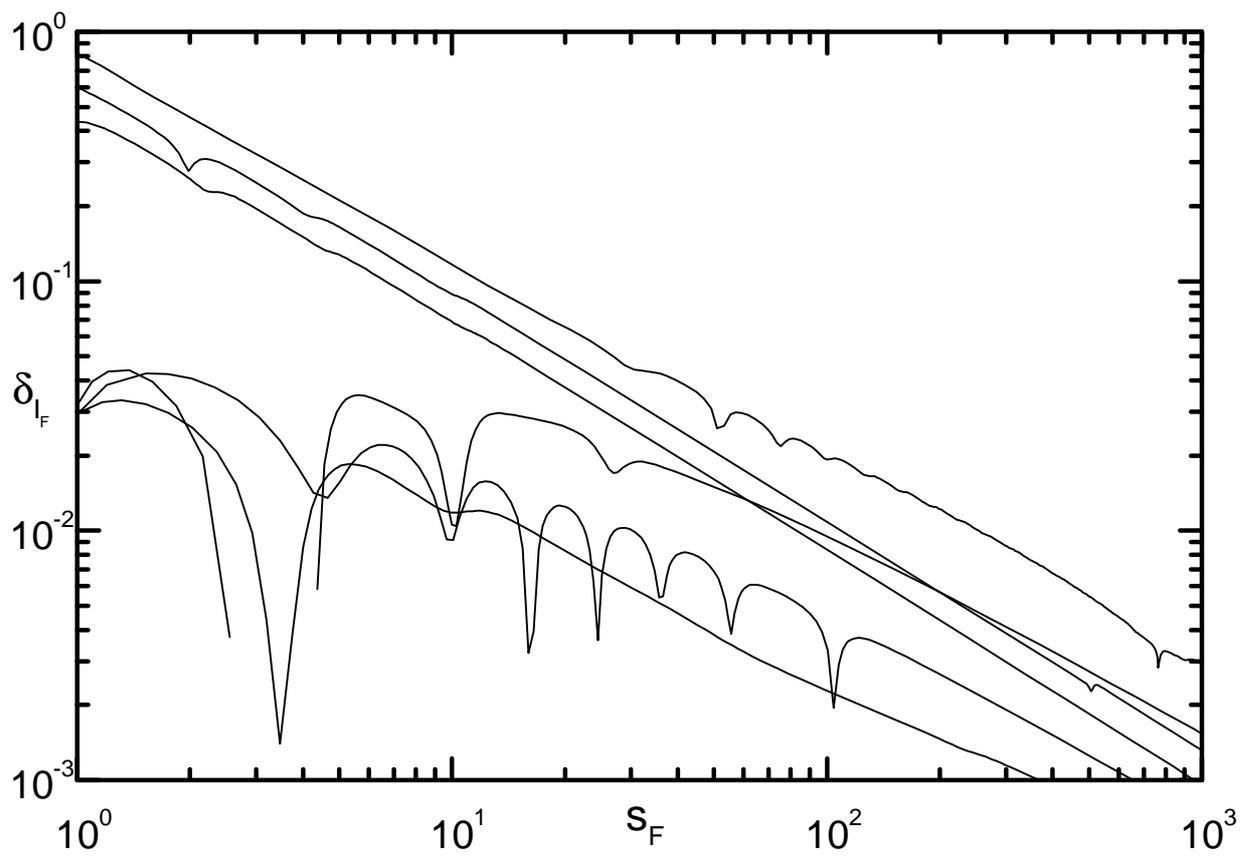
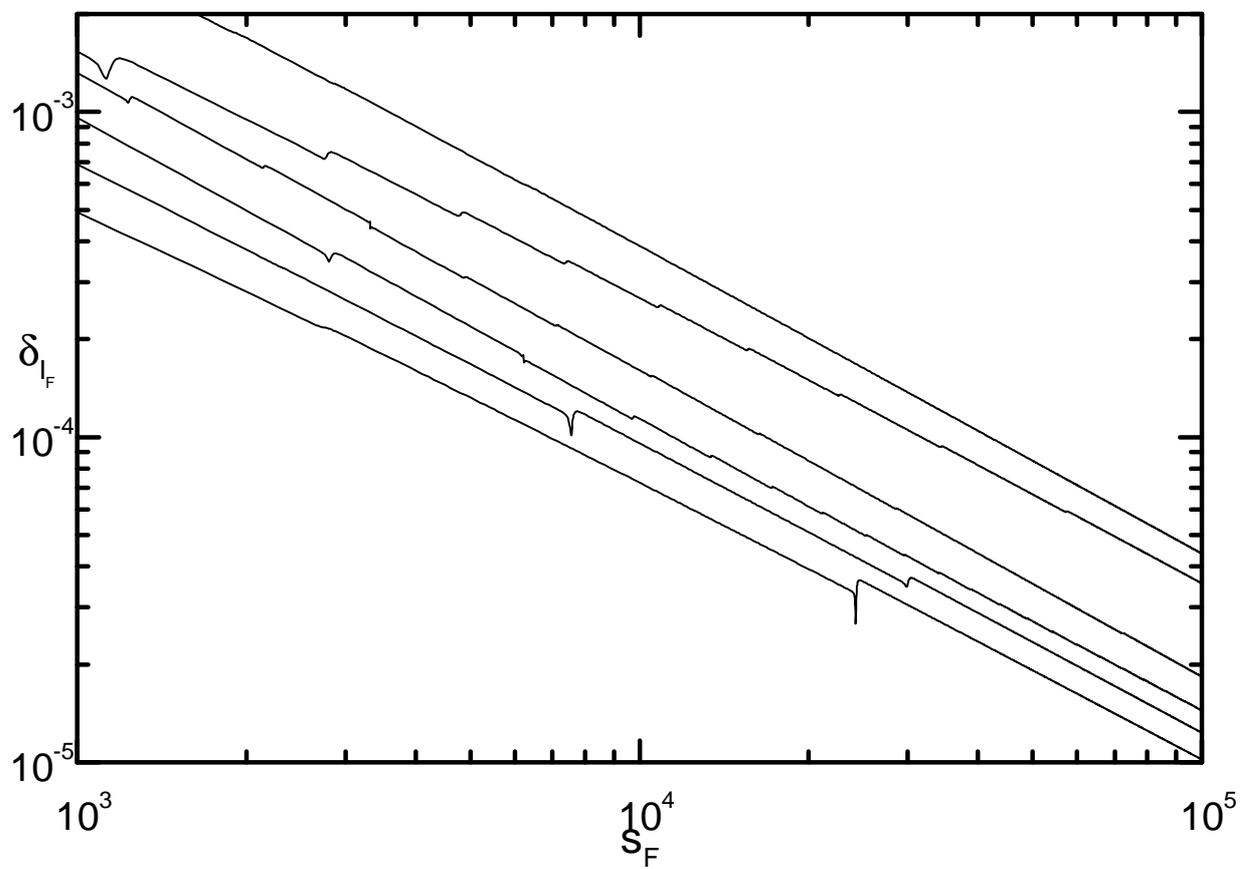
\end{document}